# Coherent Phonons in Antimony: an Undergraduate Physical Chemistry Solid-State Ultrafast Laser Spectroscopy Experiment

Ilana J. Porter[1,2], Michael W. Zuerch[1,3], Anne M. Baranger[1,4], Stephen R. Leone[1,2,5,*]

[1]Department of Chemistry, University of California, Berkeley, CA 94720, USA.

[2]Chemical Sciences Division, Lawrence Berkeley National Laboratory, Berkeley, CA 94720, USA.

[3]Materials Sciences Division, Lawrence Berkeley National Laboratory, Berkeley, CA 94720, USA.

[4]Graduate Group in Science and Mathematics Education, University of California, Berkeley, CA 94720, USA.

[5]Department of Physics, University of California, Berkeley, CA 94720, USA

**ABSTRACT**

Ultrafast laser pump-probe spectroscopy is an important and growing field of physical chemistry that allows the measurement of chemical dynamics on their natural timescales, but undergraduate laboratory courses lack examples of such spectroscopy and the interpretation of the dynamics that occur. Here we develop and implement an ultrafast pump-probe spectroscopy experiment for the undergraduate physical chemistry laboratory course at the University of California Berkeley. The goal of the experiment is to expose students to concepts in solid-state chemistry and ultrafast spectroscopy via classic coherent phonon dynamics principles developed by researchers over multiple decades. The experiment utilizes a modern high-repetition-rate 800 nm femtosecond Ti:Sapphire laser, split pulses with a variable time delay, and sensitive detection of transient reflectivity signals using the lock-in technique. The experiment involves minimal intervention from students and is therefore easy and safe to implement in the laboratory. Students first perform an intensity autocorrelation measurement on the femtosecond laser pulses to obtain their temporal duration. Then, students measure the pump-probe reflectivity of a single-crystal antimony sample to determine the period of coherent phonon oscillations initiated by an ultrafast pulse excitation, which is analyzed by fitting to a sine wave. Students who completed the experiment in-person obtained good experimental results, and students who took the course remotely due to the COVID-19 pandemic were provided with the data they would have obtained during the experiment to analyze. Evaluation of student written and oral reports reveals that the learning goals were met, and that students gained an appreciation for the field of ultrafast laser-induced chemistry.

**GRAPHICAL ABSTRACT**

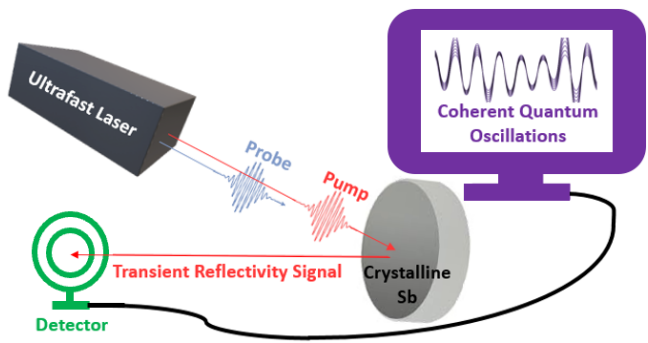

**INTRODUCTION**

Ultrafast pump-probe spectroscopy, the technique used to capture the sub-picosecond response of chemical systems to light excitation, is a major field of physical chemistry.[1] Realized through advances in pulsed laser technology and elevated by achievements from scientists such as Nobel Prize winner Prof. Ahmed H. Zewail,[2] ultrafast spectroscopy research labs are prominent at most universities and many colleges. The U.S. Department of Energy affirms that funding in this area will promote discoveries that are "critical drivers of technological innovation and economic growth", owing to its importance for both chemical and solid-state materials research and mechanisms.[3] Many careers that chemistry undergraduate students will accede to, from university professor to research scientist at a free electron laser light source to an engineer at a semiconductor company, will benefit from exposure to concepts in ultrafast spectroscopy in a laboratory setting.

Pump-probe spectroscopy is a technique in which an intense "pump" burst of light, typically a laser pulse, excites a chemical system and is followed at varying time intervals by a weaker "probe" pulse of light that samples the resulting energy state. Varying the time delay between pump and probe enables tracking chemical dynamics directly in the time domain, with the resolution typically limited by the temporal width of the pulses. Previous laboratory experiments designed to teach undergraduates this technique have focused on slow chemical reactions that can be measured with flash photolysis or a standard UV-Vis spectrometer,[4–9] with a more recent experiment involving a sub-ns laser.[10] While the equipment for these experiments is very cost effective, the nanosecond or longer reaction timescales measured are many orders of magnitude slower than the picosecond (ps) or femtosecond (fs) timescales of electronic photochemical dynamics or bond vibration recurrences that govern the quantum dynamics in chemical systems. So far, reported undergraduate experiments involving femtosecond lasers demonstrate optical physics[11-14] but not photoinitiated chemistry, which is only demonstrated experimentally to masters students.[15] Furthermore, these experiments explore gas phase and solution samples but not the response of a solid-state system to light excitation. An experiment on chemical dynamics of a condensed-phase system is at the interface of physics, materials science, and chemistry, which is a leading area of contemporary research, and exposure to this interdisciplinary field will aid graduates as they enter the workforce.

A laboratory-based ultrafast pump-probe spectroscopy experiment is not only useful as an introduction to the types of equipment and experiments utilized in this field, but the actions of performing the experiment and interacting with lab partners and teaching assistants will lead to deeper learning of the physical chemistry concepts.[16] This 'situated cognition' theory of learning posits that the entire context, physical and social, in which learning occurs affects the types of neural connections and memories that are formed.[17,18] Students engaged in an 'active' form of learning, such as discussions and guided inquiries, display improved outcomes on course assessments.[19,20] A femtosecond experiment has the advantage of emphasizing modern principles of superposition states, wave packets and coherent dynamics, relating directly to time dynamics of nonstationary states that are important in quantum mechanics.

The benefits of an undergraduate laboratory experiment utilizing ultrafast pump-probe spectroscopy are clear, but startup costs and safety concerns have prevented implementation. Now, owing to the ubiquity of femtosecond lasers in modern physical chemistry research, institutions may already be in possession of a femtosecond broadband mode-locked laser, which is the most costly item required for this experiment. An ultrafast pump-probe experiment can be designed and built on a movable optical breadboard for use in an existing laser laboratory space. Furthermore, Ti:Sapphire lasers with adequate specifications for this experiment are available for less than $45,000, including installation (see Supplemental Information). Here, a complete undergraduate laboratory experiment utilizing ultrafast pump-probe spectroscopy to probe coherent phonons in antimony is presented. The

compact setup fits on a 2 foot by 4 foot optical table and utilizes a mode-locked femtosecond oscillator (800 nm, 80 MHz repetition rate, nominally 25 fs pulse duration, model Coherent Vitara-S, on long-term loan with a reduced-cost maintenance contract from Coherent, Inc). Additional startup costs can be kept under $20,000, while including multiple layers of engineered safety controls (see Supplemental Information for parts and price list).

The ultrafast solid-state experiment is implemented in the upper division physical chemistry laboratory course (CHEM 125) at the University of California Berkeley. This course offers approximately fifteen laboratory experiments, and students choose six of them to be completed in one five-hour lab session each. Pairs of students or single students complete their experiments with minimal help from the graduate student instructor, who oversees multiple experimental options offered simultaneously. This requires that experiments are considerably robust and turn-key, which were design goals of the experiment presented here. In this experiment, students first perform an intensity autocorrelation measurement to determine the temporal duration of the laser pulses. Then, students perform an ultrafast pump-probe measurement of the transient reflectivity of single-crystal antimony to observe the $a_{1g}$ coherent phonon oscillation. To complete the entire measurement, students simply turn on the electronics, execute the LabVIEW control software, and move a single magnetically mounted mirror from one position to another while the laser beam is blocked to access the different measurements. The procedure is designed such that there is no need for the students to align the class 4 laser beam or view it with the safety enclosure opened, which increases the overall safety and repeatability of the experiment. The experiment can easily be adapted for laboratory classes with different time availability to involve increased assembly and alignment by the students. Student alignment of the intense optical pulses would require close supervision due to the potential for retinal damage upon eye exposure and skin burns.

Learning Goals

The goals of this experiment are twofold. Primarily, the aim of this experiment is to introduce students to concepts in solid-state chemistry. Students are asked to consider the ultrafast dynamics that occur in a solid-state system and their timescales, understand and observe collective atomic motions in solids (phonons), and learn about coherence and superpositions in quantum mechanical systems. The secondary goal of the experiment is to demonstrate concepts of ultrafast spectroscopy, such as femtosecond laser pulses, the pump-probe technique, the time-bandwidth relation, lock-in amplification, reflectivity of materials, nonlinear optics, data analysis, pulse measurements and autocorrelation. During the COVID-19 pandemic, students who were unable to perform this laboratory experiment in-person were provided with online lectures and walkthroughs of the experiment, as well as the data they would have obtained in the lab. In more recent semesters, students were able to complete the experiment in-person, and all of them obtained data for both the autocorrelation and pump-probe measurements with little to no instructor input. Both sets of students were able to complete written or oral reports outlining the experiment, the data analysis, and the results. Assessment of student reports and answers to discussion questions confirm that the goals of the experiment were met.

**THEORY OF THE EXPERIMENT**

One of the earliest ultrafast pump-probe spectroscopy experiments to be performed on a solid sample was the excitation and observation of the coherent $a_{1g}$ phonon in a number of opaque solid samples by the group of Dresselhaus.[21,22] In these experiments, 800 nm light from an ultrafast pulsed Ti:Sapphire mode-locked laser acts as both the pump and probe, initiating the coherent phonon motion and detecting it as modulations to the reflectivity. With modern advances in laser technology, this experiment is readily adaptable to an undergraduate laboratory experiment. Here, we present the basic theory of this phenomenon, with a more detailed description available in the manual provided to students (see Supplemental Information).

Analogous to the quantized vibrational modes of molecules, phonons are vibrational states in a crystalline solid defined by their crystallographic direction and frequency. Coherent phonon motion occurs when the phonons of a given frequency have a fixed phase relationship, i.e., the oscillations constructively interfere, which can occur when the oscillations are driven by a very fast excitation. This phenomenon had been observed using pump-probe reflectivity prior to the work of Dresselhaus.[21] in the form of Impulsive Stimulated Raman Scattering (ISRS), when a short Raman excitation rocks the atoms around their equilibrium positions. In contrast, results from Dresselhaus detected an effect in which the atoms oscillated about an offset position, measured as a jump in the total surface reflectivity, and they called the phenomenon the Displacive Excitation of Coherent Phonons (DECP).

In certain opaque solids such as Sb, Bi, Te and $Ti_2O_3$, there exists an energetically favorable breaking of the equilibrium crystalline symmetry, called a Peierls distortion, in which atoms displace along the $a_{1g}$ mode direction and lower the overall potential energy of the electrons (Figure 1(a)). This Peierls distortion creates a strong coupling between the electronic and vibrational degrees of freedom in the crystal. In these materials, very fast electronic excitation in the form of a laser pulse disrupts the electron potential energy surface, removing the energetic stabilization of the symmetry breaking and causing the atoms to change equilibrium position. Since the atoms are displaced along the $a_{1g}$ mode direction, and the electronic excitation occurs very quickly, the laser pulse excites coherent motion of the $a_{1g}$ phonons. Quantum mechanically, this is a coherent superposition of vibrational (phonon) states, initiated by the abrupt deposition of energy into the lattice. This change of equilibrium position and coherent phonon motion is the DECP phenomenon, and the resultant changes to the electronic binding energy can be observed as changes to the total reflectivity (Figure 1(b)).

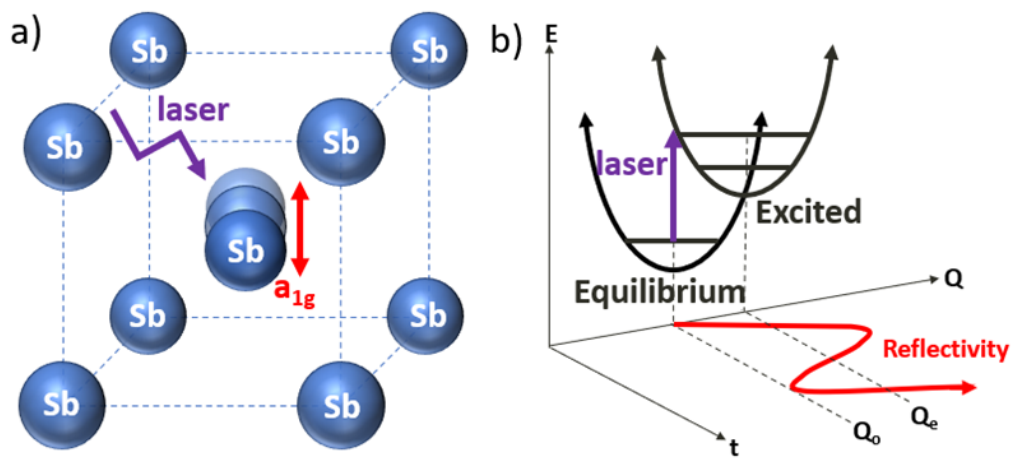

Figure 1. a) A simplified diagram of the unit cell of crystalline antimony. Each antimony atom is offset from the true center of the unit cell along the $a_{1g}$ mode direction due to the Peierls distortion. For simplicity, only the central atom is shown as offset here. After laser excitation, the central antimony atom oscillates along the $a_{1g}$ mode, crossing the true center of the unit cell. b) A schematic of the potential energy surface of crystalline antimony, with axes E for energy and Q for the reaction coordinate. On the top left is the equilibrium energy surface centered at Qo. After laser excitation, the crystal's potential energy surface changes to the excited state shown on the top right centered at Qe. On the bottom is a cartoon of the reflectivity of the antimony crystal, starting at Qo and then oscillating around the Qe position after laser excitation.

To initiate the DECP mechanism, a very short laser pulse of less than 100 fs ($10^{-13}$ s) is required, which is much faster than the response time of an electronic detector. Students are asked to infer the temporal duration of the laser pulse using an intensity autocorrelation measurement. In an intensity autocorrelation, the laser pulse is split in two and one half is delayed with respect to the other by having it travel a longer path. Students use the speed of light and distance traveled to convert to the time delay. A nonlinear optical crystal acts as the recombination medium and doubles the frequency of the light. This not only allows for easy separation and detection of the autocorrelation signal, but it also transforms the electric field measurement into an intensity measurement, from which the temporal envelope of the pulse can be directly deduced (Figure 2).[23] The path difference and, therefore, the time delay between the two half-pulses is varied to obtain the intensity autocorrelation over time. The intensity signal is then fit with a Gaussian function to obtain the temporal duration of the pulse.

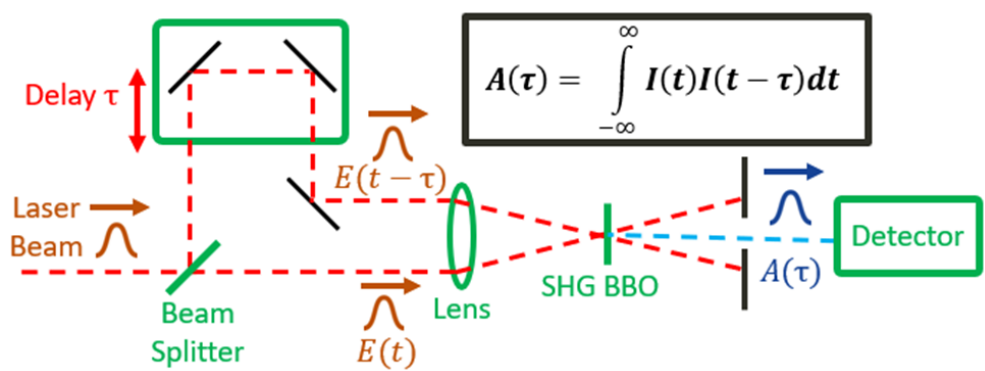

Figure 2. A schematic of an intensity autocorrelator. The laser pulse with temporal electric field E(t), indicated by the dashed red line, is split by a beam splitter. One half of the pulse is delayed by time τ, and both halves are focused by a lens into a nonlinear crystal for second harmonic generation (SHG) (beta barium borate (BBO), see Supplemental Information for details of the crystal). In the crystal, the two half-pulses are overlapped and produce a third laser pulse with double the frequency of the incoming pulses and amplitude A(τ), indicated by the dashed blue line. This third pulse is the autocorrelation of the incoming laser pulse and is separated spatially from the two incoming half-pulses to be measured on a detector (photodiode) as a function of time delay.

**HAZARDS**

The class 4 femtosecond laser used in this experiment produces extremely high peak powers, enough that even scattered light may cause permanent retinal damage. Additionally, the 800 nm near-infrared wavelengths are mostly invisible to the human eye. As engineered safety measures, a primary enclosure with a safety shutter interlock that automatically blocks the laser when the lid is open is used (Figure 3(a) and 3(b)), and a tall secondary enclosure blocks stray beams from travelling to other lab stations (Figure 3(c)). Students must wear the appropriate laser safety goggles at all times when behind the secondary enclosure, even though the experiment itself is enclosed by the primary safety box, in case of accidental movement of covers, such as in an earthquake, trip or fall.

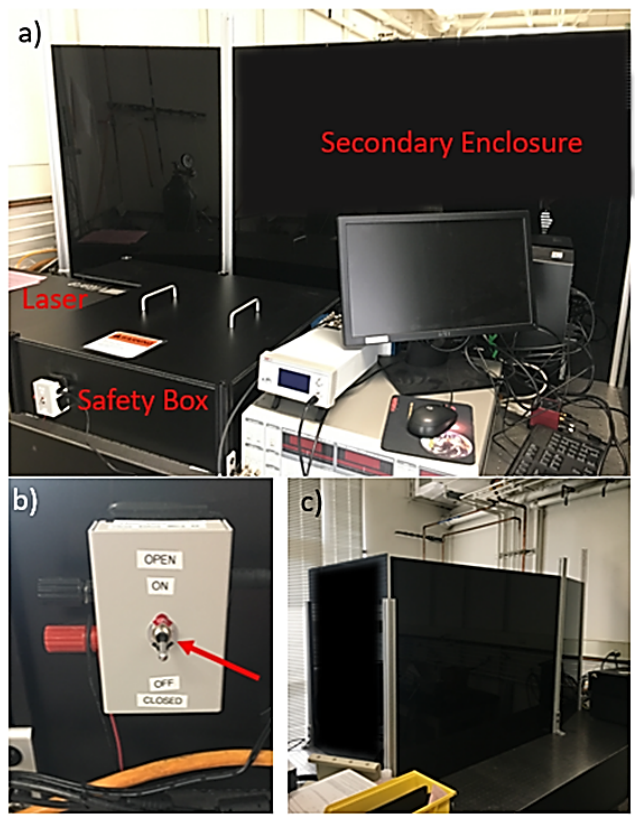

Figure 3. a) A picture of the experimental setup, surrounded on three sides by the secondary enclosure. The optical layout is entirely enclosed in the safety box on the left, which features an external safety shutter switch. On the right are the electronics controllers and computers. b) A close-up view of the safety shutter switch. c) The back side of the secondary enclosure, which prevents stray beams from exiting the area and allows other students to perform experiments in the same laboratory space.

## EXPERIMENTAL DESIGN

In-class lectures for the upper division physical chemistry laboratory class discuss concepts of lasers, absorption, photovoltaic principles and detectors, and crystal structure, and the details of this specific experiment are mainly covered in a one-hour prelab lecture given by the graduate student instructor. Students are asked to read the lab manual and answer prelab questions before attempting the lab (see Supplemental Information). Afterward, students submit individual written reports, or take an oral exam and submit written data analysis and calculations, that must include the answers to discussion questions about the experimental concepts (see Supplemental Information for grading rubrics), which were used to assess the learning goals of the experiment. Since this experiment demonstrates so many concepts in solid-state chemistry and ultrafast spectroscopy, not all of the learning goals could be included in the discussion questions, but student understanding of these concepts was demonstrated in the reports.

In this experiment, one or a pair of students perform multiple repetitions of an intensity autocorrelation measurement and a pump-probe transient reflectivity measurement within a single five-hour lab period. The measurements are designed to be fully automated by LabVIEW, and students just need to turn on the equipment and start the LabVIEW programs to control the delay stage movements and detectors. A single magnetically mounted mirror is moved from one position to another in order to transfer between the two measurements, allowing many of the same optics and electronics to be used for both experiments while maintaining safety and repeatability (Figure 4(a) and 4(b)). Students performing the lab are encouraged to open up the safety enclosure while the laser is off and study the setup so that they may identify the optics components and recognize the beam path. Here, we outline the basic components and procedures of the experiment, with more detailed steps found in the Experimental Procedures and the Guide for Instructors (see Supplemental Information).

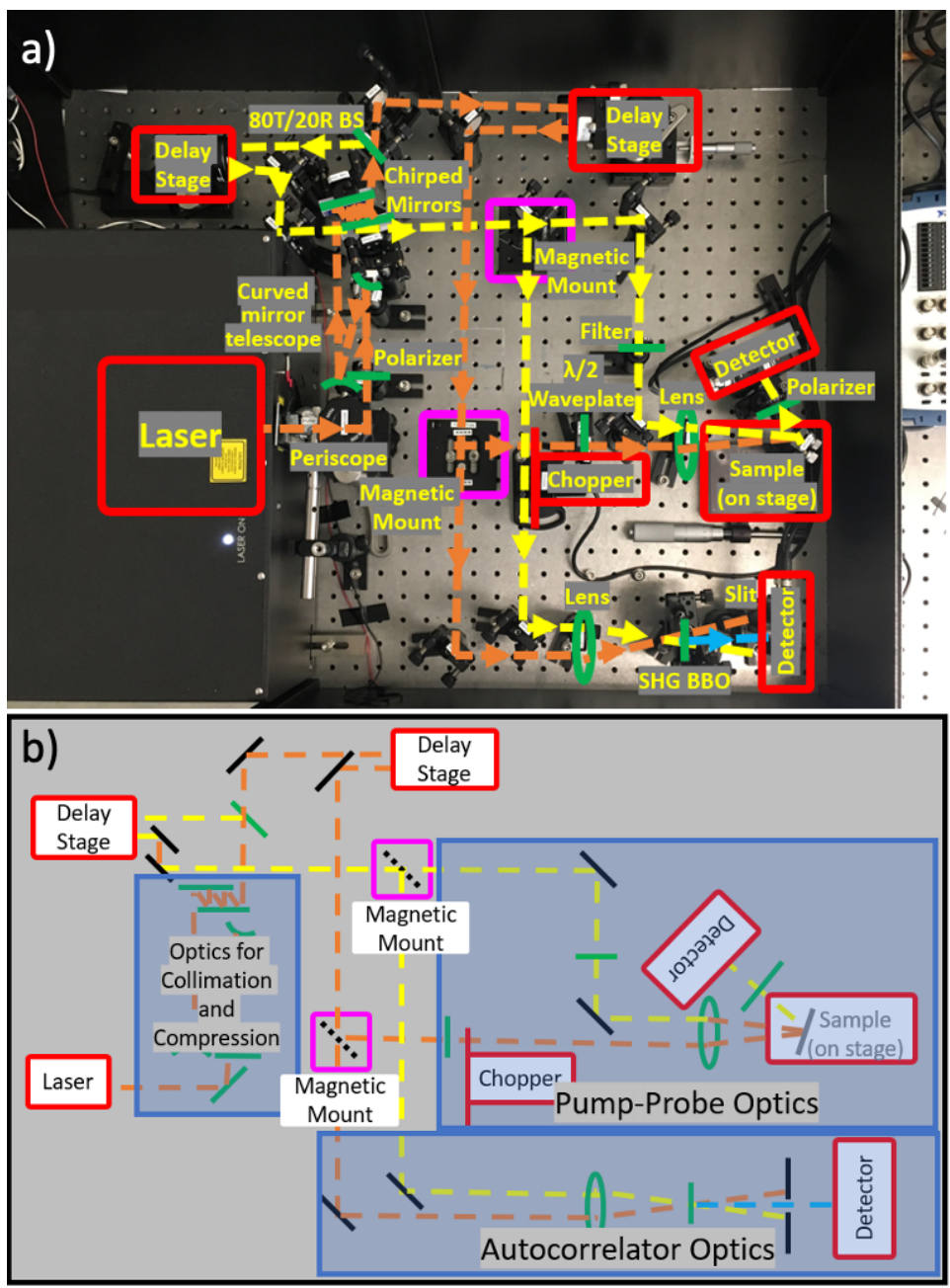

Figure 4. a) A picture of the optical table layout for the experiment. Optics other than mirrors are labeled in green. The laser beam path is the dashed orange line for the pump and the dashed yellow line for the probe, with dashed blue for the output of the autocorrelation crystal. Electronics and moving parts, like the delay stages and chopper, are shown in red, and the magnetic mounts are labeled in magenta. The two magnetic mirror mounts allow for simple and repeatable switching between the autocorrelation and pump-probe experiments. b) A schematic diagram of the optical layout, with the optics for the autocorrelator and pump-probe experiments delineated.

## EXPERIMENTAL PROCEDURE

In order to obtain adequate transient signals, pump pulses with intensity greater than approximately 1 nJ at the sample, and pump and probe pulses of less than approximately 100 fs are required. Chirped mirrors (Figure 4) are used to compensate for wavelength dispersion directly from the Coherent Vitara-S laser and from the subsequent lenses.

Both measurements utilize a 25 mm motorized translation stage (KMTS25E, Thorlabs) to delay one of the split off beams with respect to the other. The autocorrelator measurement then uses an amplified silicon photodetector (PDA36A2, Thorlabs) to measure the signal produced when the beams are overlapped in the nonlinear crystal and produce second harmonic light pulses. These devices are controlled via a custom LabVIEW program and a multifunction I/O device (USB-6212 BNC, National Instruments) as an analog-to-digital converter. To perform the measurements, students first start up the Coherent Vitara-S laser and warm up all of the electronics. Next, they unblock the laser safety shutter and start the LabVIEW program. The program performs the autocorrelation measurement by moving the delay stage and capturing the photodiode voltages, and plots the resulting intensity versus position, shown in Figure 5(a). This data is fit with a Gaussian function by the program (see Supplemental Information) and students are asked to convert from the fitted Gaussian width σ in units of the delay stage distance to the pulse full width at half maximum (FWHM) in units of time. Examples of the LabVIEW program output and data provided to the students can be found in the Supplemental Information. Students can further develop an insight into the pulse spectral bandwidth and the time duration of the pulse, based on additional information about the laser pulse spectrum provided or obtained in company literature.

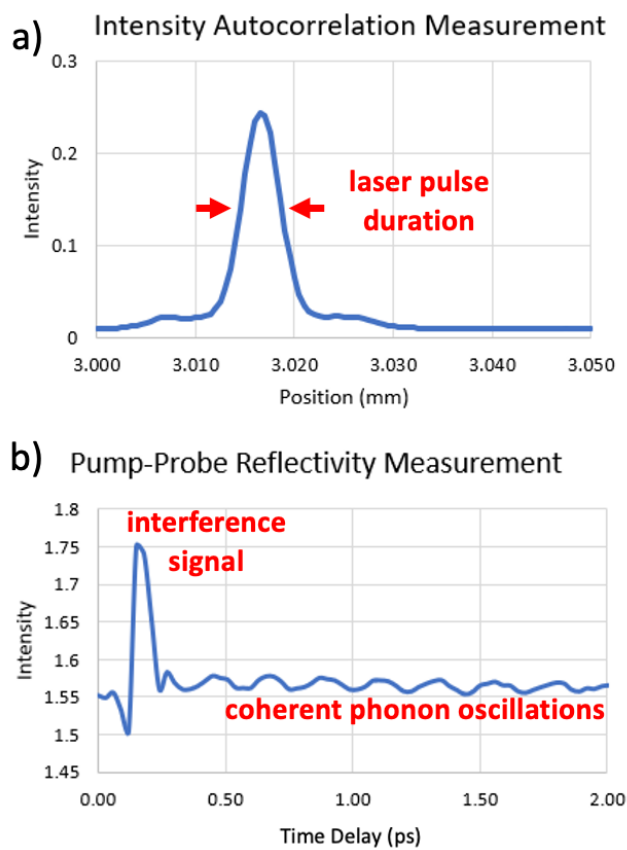

Figure 5. a) An example of the data obtained from an intensity autocorrelation measurement. The peak is the intensity autocorrelation signal, which is fit with a Gaussian for the students to determine the temporal width of the laser pulses. b) An example of the pump-probe reflectivity data obtained with the single-crystalline antimony sample. The peak near the beginning of the measurement is caused by the interference of the simultaneous pump and probe beams on the sample surface. The less-intense oscillations that continue to the end of the scan window are the coherent $a_{1g}$ phonon oscillations obtained through the reflectivity changes of the probe pulse versus time. The period of oscillation is fit by the students with sine functions and converted to a value in units of frequency.

After obtaining the autocorrelation data, students block the beam and move the magnetically mounted mirror to direct the beams into the pump-probe transient reflectivity optics. Since the sample is a polished single crystal of antimony (Goodfellow, Inc.), students do not need to perform any sample preparation. For this measurement, the same delay stage is used, and the weak transient signal is resolved with a biased silicon detector (DET36A2, Thorlabs) and then amplified with a lock-in amplifier (SR830, Stanford Research Systems). To utilize the lock-in amplifier, an optical chopper (MC2000B, Thorlabs) modulates the stream of high repetition rate pump pulses by alternately blocking and unblocking steams of pulses and is synchronized to the lock-in amplifier. This configuration was chosen because a spare lock-in amplifier was available and easy to operate, but more cost-effective alternatives can be used to amplify the transient reflectivity signal.[24] Once the students have turned on these additional components, they start the custom LabVIEW program that controls the pump-probe measurement. The LabVIEW program produces a file containing the differential reflectivity data, which is the difference between the pumped and un-pumped reflectivity, versus time in picoseconds (Figure 5(b)). Students are instructed in the manual how to fit the resulting trace to obtain the period of the oscillation. More of the sample data can be found in the Supplemental Information.

**RESULTS**

Due to the SARS-COV-2 pandemic, in-person instruction at the University of California Berkeley was halted in 2020 and students who completed this experiment in those semesters had to be provided with a video recording of the prelab lecture, a "virtual lab walkthrough" slide deck, and were sent sample data to analyze for their reports from home (see Supplemental Information). In 2021, in-person instruction resumed and all the students who chose this experiment were able to complete it in a single five-hour lab period and obtain both the autocorrelation data and pump-probe data without instructor input. Student feedback overall was positive, with a graduate student instructor stating that the experiment was "very easy to handle while still providing a good example to understand the basics of pump-probe experiments." Graduate student instructors who tested the experiment stated that it was "clearly marked for students to follow" and "straight forward to complete both the lab portion and data analysis", and that they "like how you get to go into the optical set up and move things around the table." Additionally, students commented on how "turn-key" and stable the experimental setup was, and how simple it was to obtain high-quality data just from following the manual. One instructor claimed that "students can gain experience with ultrafast spectroscopy, which is not a common opportunity for undergraduates despite its prevalence in Ph.D. research."

In total, 18 students consented to have their reports used in this study (see Supplemental Information for consent form template). Of them, 4 students who completed the experiment in Spring of 2020 (labeled 1, 2, 3, 4) and 4 who completed it in Fall of 2020 (5, 6, 7, 8) received the at-home datasets and completed the written report. The other 10 students completed the experiment in-person in Fall 2021, and 5 students (9, 10, 11, 12, 13) completed a written report while 5 students (14, 15, 16, 17, 18) completed the oral report and written data and calculations. The mean score on the written reports, out of 100, was 92 ± 4 for Spring 2020, 81 ± 11 for Fall 2020, and 91 ± 7 for Fall 2021, for an average of 88 ± 9. The mean score on the Fall 2021 oral reports was 89 ± 6, which is very close to the written report average, and yields a total average of 88 ± 8 out of 100. This is close to the average report grade for other experiments in this course. The same written report rubric was used for all three semesters (see Supplemental Information), so differences in grades between the semesters are caused by differences in report quality. The majority of points deducted were for missing details and sections, and not for misunderstanding. Since the main results of the experiment are the values for the pulse width and coherent phonon frequency, it is useful to measure the students' success in performing the analysis by the numerical results they obtained. The distribution of calculated pulse durations in femtoseconds and the distribution of calculated antimony $a_{1g}$ phonon frequencies in terahertz is shown in Table 1. The true pulse duration is approximately 22-23 fs, and the true phonon frequency is approximately 4.5 THz. Incorrect answers are marked in red. All students were able to obtain accurate phonon frequencies, and 11 of the 18 students got accurate pulse durations, with a couple of students (number 3 and 9) obtaining pulse durations that were off by a factor of 2, likely due to forgetting that the pulse travels the path length of the delay stage twice. Additional students (number 5 and 18) missed the factor of $\sqrt{2}$ to convert between the cross-correlation width and pulse duration, and another (number 8) was missing the 0.707 factor converting between the Gaussian sigma and FWHM. The accuracy of student results demonstrates an understanding of the concepts presented in this experiment.

**Table 1. Student Numerical Results**

| Student | Autocorrelation Time (fs) | Phonon Frequency (THz) | Student | Autocorrelation Time (fs) | Phonon Frequency (THz) |
|---|---|---|---|---|---|
| 1 | 22.5 ± 0.2 | 4.66 ± 0.05 | 10 | <span style="color:red">14.1 ± 0.2</span> | 4.46 ± 0.05 |
| 2 | 22.1 ± 0.3 | 4.53 ± 0.08 | 11 | 23.8336 ± 0.1527 | 4.486 ± 0.002 |
| 3 | <span style="color:red">11.0 ± 0.2</span> | 4.52 ± 0.03 | 12 | <span style="color:red">7.96 ± 0.11</span> | 4.63 ± 0.16 |
| 4 | 22.1 ± 0.3 | 4.59 ± 0.02 | 13 | 23.8 ± 0.12 | 4.43 ± 0.00046 |
| 5 | <span style="color:red">15.63 ± 6.4</span> | 4.44 ± 0.8 | 14 | 23.70 ± 0.22 | 4.44 ± 0.2 |
| 6 | 23.8 ± 0.7 | 4.52 ± 0.02 | 15 | 23.25 ± 0.95 | 4.55 ± 0.04 |
| 7 | 22.0 | 4.56 ± 0.1 | 16 | 26.0 ± 0.6 | 4.4 ± 0.1 |
| 8 | <span style="color:red">9.46 ± 0.018</span> | 4.52 ± 0.03 | 17 | 25.7 ± 0.4 | 4.55 ± 0.09 |
| 9 | <span style="color:red">12.1 ± 0.1</span> | 4.46 ± 0.09 | 18 | <span style="color:red">16.8 ± 0.2</span> | 4.478 ± 0.0140 |

To further investigate whether the learning goals were met, student answers to the discussion questions were analyzed. To determine understanding of the solid-state chemistry concepts, students are asked to explain the number of oscillations observed (Q1), to contrast the DECP mechanism utilized here with the ISRS mechanism (Q2) and to compare phonon frequencies in different elemental solids (Q3). Conceptual understanding of the ultrafast spectroscopy topics is measured by having the students calculate and explain the signal-to-noise ratio of the experiment due to fluctuations in the laser pulses (Q4) and the pulse energy and peak power of the laser (Q5). Additional exercises can be introduced to help understand the phase relationships in a coherent sum of wave functions that results in a coherent superposition of phonons, the time-bandwidth relation, or many other concepts that are raised in this experiment. On average, the 18 students received approximately 76% of the points available for answering these questions correctly, with 8% of the points lost due to missing answers. The percentage of correct answers, missing answers, math errors and other incorrect answers due to misunderstanding are shown in Table 2. Ignoring missing answers, students performed best on Q4, the ultrafast spectroscopy signal-to-noise question, making only math errors. Otherwise, the excellent reports and the high grades attest that students learned the majority of the presented concepts, even those who missed out on the in-lab experience.

**Table 2. Discussion Question Results**

| Concept | Discussion Question | % Points Correct | % Missing Answer | % Math Errors | % Other Incorrect |
|---|---|---|---|---|---|
| Solid-State Chemistry | Q1 | 71% | 10% | 0% | 19% |
| | Q2 | 69% | 14% | 0% | 17% |
| | Q3 | 88% | 0% | 4% | 8% |
| Ultrafast Spectroscopy | Q4 | 91% | 5% | 4% | 0% |
| | Q5 | 67% | 11% | 4% | 19% |
| | Total | 76% | 8% | 2% | 13% |

## CONCLUSION

An undergraduate experiment was designed to teach students concepts of ultrafast pump-probe spectroscopy and solid-state chemistry by having them perform an intensity autocorrelation measurement and a pump-probe reflectivity measurement on antimony. Aside from the femtosecond laser, startup costs for this experiment can be kept under $20,000. This experiment requires minimal student and instructor intervention and can be completed in a single 5-hour laboratory period. From

analysis of student written and oral reports, it can be concluded that the learning objectives were met and that students gained significant knowledge from this exercise. This experiment demonstrates the feasibility of including ultrafast solid-state spectroscopy in the undergraduate laboratory curriculum.

**ACKNOWLEDGMENTS**

The authors gratefully acknowledge the support of the University of California Berkeley College of Chemistry undergraduate program for funding this experiment through donor support. I.J.P. acknowledges the support of the University of California Berkeley Department of Chemistry. We personally acknowledge Dr. Alexander Paarmann at the Fritz Haber Institute of the Max Planck Society, whose fruitful discussions and advice was instrumental to improve our version of this experiment and manuscript, as well as the manual. We appreciate the support of the graduate student instructors Scott Garner, Jonathon Nessralla, Jenna Tan, Can Uzundal, Bailey Nebgen, Finn Kohrell, and Jonah Adelman as well as the undergraduate students who performed the experiment. Special thanks goes to Dante Valdez Jr., who supervises the upper division instructional support facilities at the University of California Berkeley Department of Chemistry. This experiment would not be possible without the generous support from Coherent, Inc. for the long-term loan of the Vitara-S ultrafast modelocked laser.